\documentclass[review]{elsarticle}

\usepackage{hyperref}
\usepackage{url}
\usepackage{amssymb}
\usepackage{amsmath}
\usepackage{algorithm}
\usepackage{algorithmic}
\usepackage{graphicx}
\usepackage{epstopdf}
\usepackage{epsfig} 
\usepackage{subfigure}

\makeatletter
\newcommand{\AB}{\mathbf{A}}

\newcommand{\DB}{\mathbf{D}}

\newcommand{\IB}{\mathbf{I}}

\newcommand{\XB}{\mathbf{X}}

\newcommand{\dB}{\mathbf{d}}

\newcommand{\uB}{\mathbf{u}}
\newcommand{\vB}{\mathbf{v}}
\newcommand{\wB}{\mathbf{w}}
\newcommand{\xB}{\mathbf{x}}
\newcommand{\yB}{\mathbf{y}}
\newcommand{\zB}{\mathbf{z}}

\newcommand{\betaB}{\mbox{\boldmath$\beta$\unboldmath}}

\newcommand{\argmin}{\mathop{\rm argmin}}

\makeatother

\newtheorem{theorem}{Theorem}
\newtheorem{lemma}{Lemma}
\newtheorem{definition}{Definition}

\newtheorem{corollary}{Corollary}

\newtheorem{remark}{Remark}
\newcommand*{\QEDA}{\hfill\ensuremath{\blacksquare}}

\journal{Journal of \LaTeX\ Templates}

\usepackage{numcompress}

\begin{document}

\begin{frontmatter}

\title{Stable Cosparse Recovery via $\ell_q$-analysis Optimization}

\author{Shubao Zhang, Hui Qian\corref{mycorrespondingauthor}}
\cortext[mycorrespondingauthor]{Corresponding author: Hui Qian}
\address{Department of Computer Science and Technology, Zhejiang University.}
\ead{bravemind@zju.edu.cn, qianhui@zju.edu.cn}

\author{Xiaojin Gong}
\address{Department of Information Science and Electronic Engineering, Zhejiang University.}
\ead{gongxj@zju.edu.cn}

\author{Jianying Zhou}
\address{Department of Respiratory Diseases, The First Affiliated Hospital, College of Medicine, Zhejiang University.}
\ead{zjyhz@zju.edu.cn}




\begin{abstract}
In this paper we focus on the $\ell_q$-analysis optimization problem for structured sparse learning ($0< q \leq 1$). Compared to  previous work, we establish weaker conditions for exact recovery in the noiseless case and a tighter non-asymptotic upper bound of estimate error in the noisy case. We further prove that the nonconvex $\ell_q$-analysis optimization can do recovery with a lower sample complexity and in a wider range of cosparsity than its convex counterpart. In addition, we develop an iteratively reweighted method to solve the optimization problem under the variational framework. Empirical results of preliminary computational experiments illustrate that the nonconvex method outperforms its convex counterpart.
\end{abstract}


\end{frontmatter}


\section{Introduction}

The sparse learning problem is widely studied in many areas including machine learning, statistical estimate, compressed sensing, image processing and signal processing, etc. Typically, this problem can be defined as the following linear model
\begin{equation}\label{inverse}
  \yB = {\XB}{\betaB} + \wB,
\end{equation}
where ${\betaB}\in\mathbb{R}^d$ is the  vector of regression coefficients, ${\XB}\in \mathbb{R}^{m\times d}$ is a design matrix with possibly far fewer rows than columns,  $\wB\in\mathbb{R}^m$ is a noise vector, and $\yB\in\mathbb{R}^m$ is the noisy observation. As is well known, learning with the $\ell_1$ norm (convex relaxation of the $\ell_0$ norm), such as lasso \cite{Tibshirani1996} or basis pursuit \cite{Chenabc1998}, encourages sparse estimate of $\betaB$. Recently, this approach has been extended to define structured sparsity. \citeauthor{TibshiraniTaylor2011} (\citeyear{TibshiraniTaylor2011}) proposed the generalized lasso
\begin{equation}\label{glasso}
\min_{{\betaB}}  \frac{1}{2}||\yB - \XB\betaB||_2^2 + \lambda||\DB \betaB||_1,
\end{equation}
which assumes that the parameter $\betaB$ is sparse under a linear transformation $\DB\in\mathbb{R}^{n\times d}$. An equivalent constrained version is the $\ell_1$-analysis minimization proposed by \citeauthor{Candesabc2010} (\citeyear{Candesabc2010}), i.e.,
\begin{equation}\label{analysis_l1}
\min_{{\betaB}} ||\DB \betaB||_1 ~~~\textrm{ s.t.}~~~ ||\yB - \XB\betaB||_2 \leq \epsilon ,
\end{equation}
where $\DB$ is called the analysis operator. In contrast to the lasso and basis pursuit in $\DB=\IB$,  the generalized lasso and $\ell_1$-analysis minimization make a structured sparsity assumption so that it can explore structures on the parameter. They include several well-known models as special cases, e.g., fused lasso \cite{Tibshiraniabc2005}, generalized fused lasso \cite{Viallon2014}, edge Lasso \cite{Sharpnackabc2012}, total variation (TV) minimization \cite{Rudinabc1992}, trend filtering \cite{Kimabc2009}, the LLT model \cite{Lysakerabc2003}, the inf-convolution model \cite{ChambolleLions1997}, etc. Additionally, the generalized lasso and $\ell_1$-analysis minimization have been demonstrated to be effective and even superior over the standard sparse learning in many application problems. 

The seminal work of \citeauthor{FanLi2001} (\citeyear{FanLi2001}) showed that the nonconvex sparse learning holds better properties than the convex one. Motivated by that, this paper investigates the following $\ell_q$-analysis minimization ($0<q\leq 1$) problem
\begin{equation}\label{analysis_lq}
\min_{{\betaB}} ||\DB \betaB||_q^q ~~~\textrm{ s.t.}~~~ ||\yB-{\XB \betaB}||_2 \leq \epsilon.
\end{equation}
We consider both theoretical and computational aspects. In summary, the contributions of this work are as follows:
\begin{enumerate}
 \item [$\bullet$] We establish weaker conditions for exact recovery in the noiseless case and a tighter non-asymptotic upper bound of estimate error in the noisy case. Particularly, we provide a necessary and sufficient condition guaranteeing exact recovery via the $\ell_q$-analysis minimization. To the best of our knowledge,  our work is  the first study in this issue. 
 \item [$\bullet$] We show the advantage of the nonconvex $\ell_q$-analysis minimization ($q<1$) over its convex counterpart. Specifically, the nonconvex $\ell_q$-analysis minimization can do recovery with a lower sample complexity (on the order of $q k\log (n/k)$) and in a wider range of cosparsity.
 \item [$\bullet$]  We resort to an iteratively reweighted method to solve the $\ell_q$-analysis minimization problem, which converges to a critical point.
\end{enumerate}
The numerical results are consistent with the theoretical analysis. For example, the nonconvex $\ell_q$-analysis minimization indeed can do recovery with a smaller sample size and in a wider range of cosparsity than the convex method. 

\subsection{Related Work}

\citeauthor{Candesabc2010} (\citeyear{Candesabc2010}) studied the $\ell_1$-analysis minimization problem in the setting that the observation is contaminated with stochastic noise and the analysis vector $\DB\betaB$ is approximately sparse. They provided a $\ell_2$ norm estimate error bounded by $C_0\epsilon + C_1 k^{-1/2}||\DB \betaB-(\DB \betaB)(k)||_1$ under the assumption that $\XB$ obeys the D-RIP condition $\delta_{2k}<0.08$ or $\delta_{7k}<0.6$ and $\DB$ is a Parseval tight frame \footnote{A set of vectors $\{\dB_k\}$ is a frame of $\mathbb{R}^d$ if there exist constants $0<A\leq B <\infty$ such that $ \forall \betaB\in\mathbb{R}^d, ~~ A||\betaB||_2^2 \leq ||\DB\betaB||_2^2 \leq B||\betaB||_2^2, $ where $\{\dB_k\}$ are the columns of $\DB^{T}$. When $A=B=1$, the columns of $\DB^{T}$ form a Parseval tight frame and $\DB^{T}\DB=\IB$.}. \citeauthor{Namabc2011} (\citeyear{Namabc2011}) studied the $\ell_1$-analysis minimization problem in the setting that there is no noise and the analysis vector $\DB\betaB$ is sparse. They showed that a null space property with sign pattern is necessary and sufficient to guarantee exact recovery. \citeauthor{Liuabc2012} (\citeyear{Liuabc2012}) improved the analysis in \citep{Candesabc2010}. They established an estimate error bound similar to the one in \cite{Candesabc2010} for the general frame case. And for the Parseval frame case, they provided a weaker D-RIP condition $\delta_{2k}<0.2$.

\citeauthor{TibshiraniTaylor2011} (\citeyear{TibshiraniTaylor2011}) proposed the generalized lasso and developed a LARS-like algorithm pursuing its solution path. \citeauthor{Vaiterabc2013} (\citeyear{Vaiterabc2013}) conducted a robustness analysis of the generalized lasso against noise. \citeauthor{Liuabc2013} (\citeyear{Liuabc2013}) derived an estimate error bound for the generalized lasso under the assumption that the condition number of $\DB$ is bounded. Specifically, a $\ell_2$ norm estimate error bounded by $C\lambda + ||({\XB}^T{\XB})^{-1}{\XB}^T\wB||_2$ is provided. \citeauthor{NeedellWard2013} (\citeyear{NeedellWard2013}) investigated the total variation minimization. They proved that for an image $\betaB\in\mathbb{R}^{N\times N}$, the TV minimization can stably recover it with estimate error less than $C \log(\frac{N^2}{k}) (\epsilon + ||\DB\betaB-(\DB\betaB)(k)||_1 /\sqrt{k} )$ when the sampling matrix satisfies the RIP of order $k$.

So far, all the related works discussed above consider convex optimization problem. \citeauthor{Aldroubiabc2012} (\citeyear{Aldroubiabc2012}) first studied the nonconvex $\ell_q$-analysis minimization problem (\ref{analysis_lq}). They established estimate error bound using the null space property and restricted isometry property respectively. For the Parseval frame case, they showed that the D-RIP condition $\delta_{7k}<\frac{6-3(2/3)^{2/q-2}}{6-(2/3)^{2/q-2}}$ is sufficient to guarantee stable recovery. \citeauthor{LiLin2014} (\citeyear{LiLin2014}) showed that the D-RIP condition $\delta_{2k}<0.5$ is sufficient to guarantee the success of $\ell_q$-analysis minimization. In this paper, we significantly improve the analysis of $\ell_q$-analysis minimization. For example, we provide a weaker D-RIP condition $\delta_{2k}<\frac{\sqrt{2}}{2}$. Additionally, we show the advantage of the nonconvex $\ell_q$-analysis minimization over its convex counterpart.

\section{Notation and Preliminaries}

Throughout this paper, $\mathbb{N}$ denotes the natural number. $\lfloor \cdot\rfloor$ denotes the rounding down operator. The $i$-th entry of a vector ${\betaB}$ is denoted by $\betaB_i$. The \emph{best $k$-term approximation} of a vector ${\betaB}\in\mathbb{R}^d$ is obtained by setting its $d-k$ insignificant components to zero and denoted by ${\betaB}(k)$. The $\ell_q$ norm of a vector ${\betaB}\in\mathbb{R}^d$ is defined as $||{\betaB}||_q = (\sum_{i=1}^d |\beta_i|^q)^{1/q}$ \footnote{$||\betaB||_q$ for $0<q<1$ is not a norm, but $d(\uB,\vB)=||\uB-\vB||_q^q$ for $\uB,\vB\in \mathbb{R}^d$ is a metric.} for $0<q<\infty$. When $q$ tends to zero, $||{\betaB}||_q^q$ is the $\ell_0$ norm $||\betaB||_0$ used to measure the \emph{sparsity} of $\betaB$.  $\sigma_{k}(\betaB)_q = \inf_{\zB\in\{\zB\in\mathbb{R}^d: ||\zB||_0\leq k\}} ||\betaB-\zB||_q$ denotes the best $k$-term approximation error of $\betaB$ with the $\ell_q$ norm. The $i$-th row of a matrix ${\DB}$ is denoted by $\DB_{i.}$. $\sigma_{max}(\DB)$ and $\sigma_{min}(\DB)$ denote  the maximal and minimal nonzero singular value of $\DB$, respectively. Let $\kappa=\frac{\sigma_{max}(\DB)}{\sigma_{min}(\DB)}$, and  $\textrm{Null} (\XB)$ denote the null space of $\XB$.

Now we introduce some concepts related to the $\ell_q$-analysis minimization problem (\ref{analysis_lq}). The number of zeros in the analysis vector $\DB \betaB$ is refered to as \emph{cosparsity} \citep{Namabc2011}, and defined as $l:= n-||\DB \betaB||_0$. Such a vector ${\betaB}$ is said to be $l$-cosparse. The $support$ of a vector ${\betaB}$ is the collection of indices of nonzeros in the vector, denoted by $T:=\{i : \beta_i\neq0\}$. $T^c$ denotes the complement of $T$. The indices of zeros in the analysis vector $\DB \betaB$ is defined as the $cosupport$ of ${\betaB}$, and denoted  by $\Lambda:=\{j : \langle \DB_{j.},{\betaB} \rangle=0\}$. The submatrix $\DB_{T}$ is constructed by replacing the rows of $\DB$ corresponding to $T^c$ by zero rows. Denote $\DB_T\betaB=(\DB\betaB)_T$. Based on these concepts, we can see that a $l$-cosparse vector $\betaB$ lies in the subspace $\mathcal{W}_{\Lambda} :=\{{\betaB}:\DB_{\Lambda}{\betaB}=\boldsymbol{0}, |\Lambda|= l\} = \textrm{Null}(\DB_{\Lambda})$. Here $|\Lambda|$ is the cardinality of $\Lambda$.

\section{Main Results}
We begin with introducing the notion of $\mathcal{A}$-restricted q-isometry property, which is a natural generalization of restricted q-isometry property to any linear subspace \cite{ChartrandStaneva2008}.
\begin{definition}
[{$\mathcal{A}$-restricted q-isometry property}] A matrix $\boldsymbol{\Phi}\in\mathbb{R}^{m\times d}$ obeys the $\mathcal{A}$-restricted q-isometry property with constant $\delta_{\mathcal{A}}$ over any subset $\mathcal{A}\in\mathbb{R}^d$, if $\delta_{\mathcal{A}}$ is the smallest quantity satisfying
\[ (1-\delta_{\mathcal{A}})||{\vB}||_2^q \leq ||\boldsymbol{\Phi}{\vB}||_q^q \leq (1+\delta_{\mathcal{A}})||{\vB}||_2^q \]
for all ${\vB}\in\mathcal{A}$.
\end{definition}

Usually, the observation is contaminated with stochastic noise ($\epsilon\neq 0$) and the analysis vector $\DB\betaB^{*}$ is approximately sparse. This is of great interest for many applications. Our goal is to provide estimate error bound between the population parameter $\betaB^{*}$ and the minimizer $\hat{\betaB}$ of the $\ell_q$-analysis minimization (\ref{analysis_lq}).

\begin{theorem}\label{thm:1}
Assume that the analysis dictionary $\DB$ has full column rank. Let $(\DB\xB)_S$ be the best $S$-term approximation of $\DB\xB$ and $\rho = M/S (\rho> 4)$. If the condition number of $\DB$ satisfies $\kappa < (\rho^{1-q/2}-1)^{1/q}$ and the following condition holds                                                                                                                                                                                                                                                              \begin{align}\label{rip_cond}
\delta_M + (\kappa^{-2q}\rho^{1-q/2} - \kappa^{-q}) \delta_{S+M} < \kappa^{-2q}\rho^{1-q/2} - \kappa^{-q} -1,
\end{align}
then the solution $\bar{\xB}$ of the $\ell_q$-analysis minimization problem satisfies 
\begin{align*}
& ||\xB-\bar{\xB}||_2^q  \leq C_1 arepsilon^q + C_2 \frac{||\DB\xB-(\DB\xB)_S||_q^q }{S^{1-q/2}}   \\
& || \DB\xB- \DB\bar{\xB} ||_q^q \leq  C_3 \varepsilon^{q} +  C_4 ||\DB\xB-(\DB\xB)_S||_q^q
\end{align*}
with 
\begin{align*}
 &C_1 = \frac{2^q\kappa^q m^{1-q/2}} {(1-\kappa^{q}\rho^{q/2-1})[(1-\delta_{S+M})- (\kappa^q/(\kappa^{-q}\rho^{1-q/2}-1)) (1+\delta_M)]} , \\
 &C_2 = \frac{2\sigma_{min}^{-q}(\DB) \rho^{q/2-1} (1-\delta_{S+M}) }{(1-\kappa^{q}\rho^{q/2-1})[(1-\delta_{S+M})- (\kappa^q/(\kappa^{-q}\rho^{1-q/2}-1)) (1+\delta_M)]},  \\
 &C_3 = \frac{2^{q+1} m^{1-q/2} S^{1-q/2} \kappa^q}{\sigma_{min}^{q}(\DB^{+}) (1-\kappa^q\rho^{q/2-1}) [(1-\delta_{S+M})- (\kappa^q/(\kappa^{-q}\rho^{1-q/2}-1)) (1+\delta_M)]} , \\
 &C_4 =  \left[  \frac{4\kappa^{2q} \rho^{q/2-1} (1+\delta_M)}{(1-\kappa^q\rho^{q/2-1}) [(\kappa^{-q} \rho^{1-q/2} -1) (1-\delta_{S+M})- \kappa^q (1+\delta_M)]}  +\frac{4\kappa^q \rho^{q/2-1}}{1-\kappa^q\rho^{q/2-1}} +2 \right]  .
\end{align*}
\end{theorem}

This theorem says that although the $\ell_q$-analysis minimization ($q<1$) is a nonconvex optimization problem with many local minimas, one still can find the global optimum under the condition (\ref{rip_cond}) in the case that there is no noise and $\DB\betaB$ is exactly sparse. As pointed out by
\citeauthor{BlanchardThompson2009} (\citeyear{BlanchardThompson2009}), the higher-order RIP condition, just as (\ref{rip_cond}), is easier to be satisfied by a larger subset of matrix ensemble such as Gaussian random matrices. Thus, our result is meaningful both theoretically and practically. 

A slightly stronger condition than (\ref{rip_cond}) is 
\begin{align}\label{rip_scond}
 \delta_{(\rho+1)S} < \frac{\kappa^{-2q}\rho^{1-q/2} - \kappa^{-q} -1}{\kappa^{-2q}\rho^{1-q/2} - \kappa^{-q} +1}.
\end{align}
It is easy to verify that the right-hand side of the condition (\ref{rip_cond}) is monotonically decreasing with respect to $q\in(0,1]$ when $t\geq 1$. Therefore, in terms of the $\mathcal{A}$-RIP constant $\delta_{(t^q+1)k}$ with order higher than $2k$, the condition (\ref{rip_cond}) is relaxed if we use the $\ell_q$-analysis minimization ($q<1$) instead of the $\ell_1$-analysis minimization. A resulted benefit is that the nonconvex $\ell_q$-analysis minimization allows more sampling matrices to be used than its convex counterpart in compressed sensing.  Given a $\rho$, a larger condition number $\kappa$ will make the condition (\ref{rip_cond}) more restrictive, because the value of the inequality's right-hand side becomes smaller. In other words, an analysis operator with a too large condition number could let the $\ell_q$-analysis  minimization fail to do recovery. This provides hints on the evaluation of the analysis operator. For example, it is reasonable to choose a tight frame as the analysis operator in some signal processing applications. When $q$ tends to zero, the following result is straightforward.

\begin{corollary}\label{cor:2}
Let $\betaB\in\mathbb{R}^d$, $\yB=\XB\betaB$, and $||\DB\betaB||_0=k$. Assuming that $\delta_{(\rho+1)S} < \frac{\rho -2}{\rho}$, then there is some small enough $q>0$ such that the minimizer of the $\ell_q$-analysis minimization problem (\ref{analysis_lq}) with $\epsilon=0$ is exactly $\betaB$.
\end{corollary}

The $\ell_2$ error bound shows that the $\ell_q$-analysis optimization can stably recover the approximately cosparse vector in presence of noise. Again, we can see that a too ill-conditioned analysis operator leads to bad performance. Additionally, a $\ell_q$ error bound of the difference between $\betaB^{*}$ and $\hat{\betaB}$ in the analysis domain is provided, which will be used to show the advantage of the $\ell_q$-analysis minimization in the next subsection.

The linear model (\ref{inverse}) with Gaussian noise is of particular interest in machine learning and signal processing. 
Lemma 1 in \citep{CaiTabc2009} shows that the noise vector $\wB\sim N(0,\sigma^2 \IB)$ is upper bounded  by $\sigma \sqrt{m+2\sqrt{m\log m}}$ with probability at least $1-\frac{1}{m}$. The following result is thus evident.
\begin{corollary}
 If the matrix $\XB\in\mathbb{R}^{m\times d}$ satisfies the $\mathcal{A}$-RIP condition (\ref{noisy_cond}) and the noise vector $\wB\sim N(0,\sigma^2 \IB)$, then the minimizer $\hat{\betaB}$ of (\ref{analysis_lq}) satisfies
\begin{align*}
 ||\hat{\betaB} - \betaB^{*} ||_2 \leq  \frac{2 c_1 }{\sigma_{min}(\boldsymbol{D}) } \sigma \sqrt{m+2\sqrt{m\log m}}  + \frac{ 2^{1/q}  ( 2c_2 + 1) }{\sigma_{min}(\boldsymbol{D}) } \frac{ \sigma_{k}(\DB\betaB)_q }{k^{1/q-1/2}}
\end{align*}
 with probability at least $1-\frac{1}{m}$.
\end{corollary}

\section{Benefits of Nonconvex $\ell_q$-analysis Minimization}
The advantage of the nonconvex $\ell_q$-analysis minimization over its convex counterpart is two-fold: the nonconvex approach can do recovery with a lower sample complexity and in a wider range of cosparsity. 

\subsection{Sample Complexity}
In this section, we will determine how many random Gaussian measurements are needed for the $\mathcal{A}_q$-RIP to be satisfied with high probability. Our result is an extension of \cite{ChartrandStaneva2008}, and we begin with the following useful lemma in \cite{ChartrandStaneva2008}.
\begin{lemma} \label{RIP_q} [1, Lemma 3.3.]
 Let $0<q\leq 1$, $\xB\in\mathbb{R}^n$, and $\AB \in\mathbb{R}^{m\times n}$ with entries drawn from a Gaussian distribution, i.e. $\AB _{ij}\sim N(0,\sigma^2)$. Let $\delta_{\mathcal{A}}>0$ and choose $\eta,\epsilon>0$ such that $\frac{\eta+\epsilon^q}{1-\epsilon^q}\leq \delta_{\mathcal{A}}$. Let $\mu_q=\sigma^q 2^{q/2} \Gamma(\frac{q+1}{2})/\sqrt{\pi}$. Then 
 \begin{align}
  m\mu_q (1-\delta_{\mathcal{A}}) ||\xB||_2^q \leq ||\AB \xB||_q^q \leq  m\mu_q (1+\delta_{\mathcal{A}}) ||\xB||_2^q
 \end{align}
holds uniformly for $\xB\in\mathbb{R}^n$ with probability exceeding $1-P_{m,q}(\eta) = 1-2 (1+\frac{2}{\epsilon})^n e^{-\frac{\eta^2 m}{2qc_q^2}}$, where
\[
 c_q \leq (31/40)^{1/4} [1.13 + \sqrt{q} (\frac{\Gamma(\frac{q+1}{2})}{\sqrt{\pi}})^{-1/q}].
\]
\end{lemma}

Now we generalize the above proposition to any linear subspace $\mathcal{A}\in\mathbb{R}^n$ with dimension $d$. Let $\boldsymbol{U}\in\mathbb{R}^{n\times d}$ denote the basis of the subspace $\mathcal{A}$ and normalize it such that $||\boldsymbol{U}_{.j}||_2=1$. Then any $\xB\in\mathcal{A}$ can be expressed as $\xB = \boldsymbol{U}\boldsymbol{z}$ with unique coefficients $\boldsymbol{z}\in\mathbb{R}^d$. The following lemma shows that the gaussian random matrix $\AB $ still satisfies the $\mathcal{A}_q$-RIP over a linear subspace. 

\begin{lemma} \label{RIP_union} 
 Let $0<q\leq 1$, the linear subspace $\mathcal{A}\in\mathbb{R}^n$ with dimension $d$, and $\AB \in\mathbb{R}^{m\times n}$ with entries drawn from a Gaussian distribution, i.e. $\AB _{ij}\sim N(0,\sigma^2)$. Let $\delta_{\mathcal{A}}>0$ and choose $\eta,\epsilon>0$ such that $\frac{\eta+\epsilon^q}{1-\epsilon^q}\leq \delta_{\mathcal{A}}$. Let $\mu_q=\sigma^q 2^{q/2} \Gamma(\frac{q+1}{2})/\sqrt{\pi}$. Then 
 \begin{align}\label{rip_subspace}
  m\mu_q (1-\eta) ||\xB||_2^q \leq ||\AB \xB||_q^q \leq  m\mu_q (1+\eta) ||\xB||_2^q
 \end{align}
holds uniformly for $\xB\in\mathcal{A}$ with probability exceeding $1-2 (1+\frac{2}{\epsilon})^{d} e^{-\frac{\eta^2 m}{2qc_q^2}}$. 
\end{lemma}
\begin{proof}
First, we show that $\AB \boldsymbol{U}$ is still a random guassian matrix with independent entries. 
 \[
  (\AB \boldsymbol{U})_{ij} = \sum_{k=1}^n \AB _{ik} \boldsymbol{U}_{kj} \sim N(0,\sigma^2).
 \]
 \begin{align*}
  cov[(\AB \boldsymbol{U})_{ij},(\AB \boldsymbol{U})_{gl}] = E[ (\AB _{i.} \boldsymbol{U}_{.j})\cdot(\AB _{g.} \boldsymbol{U}_{.l}) ] = E[ \boldsymbol{U}_{.j}^T \AB _{i.}^T \AB _{g.} \boldsymbol{U}_{.l} ] =  \boldsymbol{U}_{.j}^T  E[ \AB _{i.}^T \AB _{g.} ] \boldsymbol{U}_{.l} 
 \end{align*}
Thus we have 
\begin{align*}
 cov[(\AB \boldsymbol{U})_{ij},(\AB \boldsymbol{U})_{gl}] = \left\{  \begin{array}{ll}
  1  &  \textrm{if } i=g, j=l\\
  0  &  \textrm{otherwise}
 \end{array} \right..
\end{align*}

According to Lemma \ref{RIP_q}, the following 
 \begin{align*}
  m\mu_q (1-\eta) ||\boldsymbol{z}||_2^q \leq ||\AB \boldsymbol{U}\boldsymbol{z}||_q^q \leq  m\mu_q (1+\eta) ||\boldsymbol{z}||_2^q
 \end{align*}
holds uniformly for $\boldsymbol{z}\in\mathbb{R}^d$ with probability exceeding $1-2 (1+\frac{2}{\epsilon})^{d} e^{-\frac{\eta^2 m}{2qc_q^2}}$. And the above inequality can be rewritten as
 \begin{align*}
  m\mu_q (1-\eta) ||\xB||_2^q \leq ||\AB \xB||_q^q \leq  m\mu_q (1+\eta) ||\xB||_2^q
 \end{align*}
for any $\xB\in\mathcal{A}$. \QEDA

\end{proof}

The following theorem provides the number of random Gaussian measurements are needed for the $\mathcal{A}_q$-RIP to be satisfied with high probability.
\begin{theorem}\label{thm:2}
 Let $\AB $ be a gaussian random matrix with i.i.d entries, i.e. $\AB _{ij}\sim N(0,\sigma^2)$, and $\mathcal{A}$ be a union of $L$ linear subspaces with dimension $d$. If
 \begin{align}
 m &\geq \frac{4qc_q^2}{\eta^2} [\ln (L) + d\ln(1+2/\epsilon) ]
 \end{align}
with $\eta,\epsilon>0$ satisfying $\frac{\eta+\epsilon^q}{1-\epsilon^q}\leq \delta_{\mathcal{A}}$, then the matrix $\AB$ satisfies the $\mathcal{A}_q$-RIP over the union of linear subspaces $\mathcal{A}$ with probability at least $1-2e^{-c(q)m}$ where $c(q)=\frac{\eta^2}{4q c_q^2}$.
\end{theorem}
\begin{proof}
Let $\mu_q=\sigma^q 2^{q/2} \Gamma(\frac{q+1}{2})/\sqrt{\pi}$, $\delta_{\mathcal{A}}>0$ and choose $\eta,\epsilon>0$ such that $\frac{\eta+\epsilon^q}{1-\epsilon^q}\leq \delta_{\mathcal{A}}$. According to Lemma \ref{RIP_union}, the matrix $\AB$ will fail to satisfy (\ref{rip_subspace}) over the union of $L$ linear subspace $\mathcal{A}$ (with dimension $d$) with probability
\begin{align*}
 \leq 2L (1+\frac{2}{\epsilon})^{d} e^{-\frac{\eta^2 m}{2qc_q^2}}
             &= 2 e^{\ln (L) +d\ln(1+2/\epsilon) - \frac{\eta^2 m}{2qc_q^2}} .
\end{align*}
It suffices to show that the righthand side of the above quantity can be bounded by $2e^{ - \frac{\eta^2 m}{4qc_q^2} }$, i.e.
\[
 \frac{\eta^2 m}{4qc_q^2}  \geq  \ln (L) + d\ln(1+2/\epsilon).
\]
It is equal to show that 
\begin{align*}
 m &\geq \frac{4qc_q^2}{\eta^2} [\ln (L) + d\ln(1+2/\epsilon) ] .
\end{align*}
Therefore, Theorem \ref{thm:2} holds with probability exceeding $1-2e^{-c(q)m}$ where $c(q)=\frac{\eta^2}{4q c_q^2}$.  \QEDA
\end{proof}

\begin{theorem}\label{thm:3}
 Let $\AB $ be a gaussian random matrix with i.i.d entries, i.e. $\AB _{ij}\sim N(0,\sigma^2)$. Then there exists constants $C_1(q)$ and $C_2(q)$ such that if $m\geq C_1(q) s + q C_2(q) s\log (n/s)$ for $0<q\leq 1$, the following is true with probability exceeding $1-2 e^{-c(q)m}$: for any $l$-cosparse signal $\xB\in\mathbb{R}^n$, $\xB$ is the unique solution of problem (1).
\end{theorem}
\begin{proof}
 The proof shares the same procedure of proof of Theorem $1.1$ in [2]. Let $M=(\rho+1)S$ and $b=\kappa^{-2q}\rho^{1-q/2} - \kappa^{-q}$. Theorem 1 states that the following slightly stronger condition  
 \[
 \delta_{(\rho+1)S} < \frac{\kappa^{-2q}\rho^{1-q/2} - \kappa^{-q} -1}{\kappa^{-2q}\rho^{1-q/2} - \kappa^{-q} +1} = \frac{b-1}{b+1}    
 \]
 guarantees the uniqueness of solution in the case that there is no noise and $\DB\xB$ is $S$-sparse. We choose $\eta = r(b-1)/(b+1)$ for $r\in(0,1)$ and $\epsilon^q = (1+r)(b-1)/2b$ to satisfy
 \[
  \frac{\eta+\epsilon^q}{1-\epsilon^q} \leq  \delta_{(\rho+1)S} \leq \frac{b-1}{b+1}.
 \]
According to Lemma \ref{RIP_union}, the matrix $\AB$ will fail to satisfy (\ref{rip_subspace}) over the linear subspace $\mathcal{A}=\{ \vB : \vB=\DB^{+}\xB,~ ||\xB||_0\leq M, ~\xB\in\mathbb{R}^d \}$ with probability
\begin{align*}
 \leq 2\left( \begin{array}{l}
              d\\
              M
             \end{array} \right)  (1+\frac{2}{\epsilon})^{M} e^{-\frac{\eta^2 m}{2qc_q^2}}
             &= 2 (e d/M)^M (1+\frac{2}{\epsilon})^{M} e^{-\frac{\eta^2 m}{2qc_q^2}}  \\
             &= 2 e^{M\ln(e d/M) +M\ln(1+2/\epsilon) - \frac{\eta^2 m}{2qc_q^2}} .
\end{align*}
It suffices to show that the righthand side of the above quantity can be bounded by $2 e^{ - \frac{\eta^2 m}{4qc_q^2} }$, i.e.
\[
 \frac{\eta^2 m}{4qc_q^2}  \geq  M\ln(e d/M) +M\ln(1+2/\epsilon).
\]
It is equal to show that 
\begin{align*}
 m &\geq \frac{4qc_q^2}{\eta^2} [ M\ln(e d/M) +M\ln(1+2/\epsilon) ] \\
 &= \frac{4qc_q^2}{\eta^2} [ (\rho+1)S\ln(e d/(\rho+1)S) +(\rho+1)S \ln(1+2/\epsilon) ] \\
 &= \frac{4qc_q^2}{\eta^2} \{ (\rho+1)S [\ln(d/S)+ \ln(e/(\rho+1))] +(\rho+1)S (\ln2 +1/q \ln(1/\epsilon^q)) \} \\
 &= \frac{4qc_q^2}{\eta^2} (\rho+1)S \ln(d/S)+ \frac{4qc_q^2}{\eta^2} (\rho+1) ( \ln\frac{2e}{\rho+1} + \frac{1}{q} \ln\frac{2b}{(1+r)(b-1)} ) S.  \\
\end{align*}
Therefore, Theorem \ref{thm:3} holds with probability exceeding $1-2e^{-c(q)m}$ where $c(q)=\frac{\eta^2}{4q c_q^2}$.  \QEDA
\end{proof}

The following theorem is a natural extension of Theorem 2.7 of \cite{Fourcartabc2010} in which $\DB=\IB$.
\begin{theorem}
 Let $m,n,k\in\mathbb{N}$ with $m,k<n$. Suppose that a matrix $\XB\in\mathbb{R}^{m\times d}$, a linear operator $\DB\in\mathbb{R}^{n\times d}$ and a decoder $\bigtriangleup :\mathbb{R}^m\rightarrow \mathbb{R}^d$ solving $\yB=\XB\betaB$ satisfy for all $\betaB\in\mathbb{R}^d$,
 \[
  ||\DB\betaB- \bigtriangleup (\XB\betaB)||_q^q \leq C \sigma_{k}(\DB\betaB)_q^q
 \]
with some constant $C>0$ and some $q$ satisfying $0<q\leq 1$. Then the minimal number of samples $m$ obeys
\begin{equation*}
 m \geq C_1 q k \log (n/4k)
\end{equation*}
with $k=||\DB\betaB||_0$ and $C_1=1/(2\log(2C+3))$.
\end{theorem}
Define the decoder $\bigtriangleup(\XB\betaB) := \DB \bigtriangleup_0(\yB)$ with $\bigtriangleup_0(\yB) := \argmin_{\betaB,\yB=\XB\betaB} ||\DB\betaB||_q^q$. Combining with the $\ell_q$ error bound in Theorem \ref{stable_rip}, we attain the following result. 
\begin{corollary} \label{cor:5}
 To stably recover the population parameter $\betaB^{*}$, the minimal number of samples $m$ for the $\ell_q$-analysis minimization must obey
 \begin{equation}
 m \geq C_2 q k \log (n/4k),
\end{equation}
where $k=||\DB\betaB^{*}||_0$ and $C_2=1/(2\log(8c_2^q+7))$ ($c_2$ is the constant in Theorem \ref{stable_rip}).
\end{corollary}
\begin{remark}
\rm In our analysis of the estimate error above, we used the $\mathcal{A}$-RIP over the set $\mathcal{A}=\{{\DB\vB}: ||\vB||_0 \leq (t^q+1)k \}$, i.e., the D-RIP. As pointed out by \citeauthor{Candesabc2010} (\citeyear{Candesabc2010}), random matrices with Gaussian, subgaussian, or Bernoulli entries satisfy the D-RIP with sample complexity on the order of $k\log (n/k)$. It is consistent with Corollary~\ref{cor:5} in the case $q=1$. However, we see that the $\ell_q$-analysis minimization can have a lower sample complexity than the $\ell_1$-analysis minimization. Additionally, to guarantee the uniqueness of a $l$-cosparse solution of $\ell_0$-analysis minimization, the minimal number of samples required should satisfy the following condition:
\begin{equation*} \label{sampling_lb}
   m\geq 2\cdot \max_{|\Lambda|\geq l} \textrm{dim}(\mathcal{W}_{\Lambda}),
\end{equation*}
where $\mathcal{W}_{\Lambda}=\textrm{Null}(\DB_{\Lambda})$. Please refer to \cite{Namabc2011} for more details. Therefore, the sample complexity of $\ell_q$-analysis minimization is lower bounded by $2\cdot \max_{|\Lambda|\geq l} \textrm{dim}(\mathcal{W}_{\Lambda})$.
\end{remark}

\subsection{Range of Cosparsity}
The condition (\ref{rip_cond}) guarantees that cosparse vectors can be exactly recovered via the $\ell_q$-analysis minimization. Define $S_q$ ($0<q\leq1$) as the largest value of the sparsity $S\in\mathbb{N}$ of the analysis vector $\DB\betaB$ such that the condition (\ref{rip_cond}) holds for some $t^q\in\frac{1}{S}\mathbb{N}$. The following theorem indicates the relationship between $S_q$ with $q<1$ and $S_1$ with $q=1$.
\begin{theorem} \label{thm:5}
Suppose that there exist $S_1\in\mathbb{N}$ and $t\in\frac{1}{S_1}\mathbb{N}$ such that
 \[
  \delta_{(t+1)S_1} <\frac{\rho(1-\kappa^4) + \kappa^2 \sqrt{4\rho+1}} { \rho (\kappa^2+1)^2 + \kappa^2 }
\]
with $\rho= \frac{1}{4} t^{-1}$.
Then there exist $S_q\in\mathbb{N}$ and $l^q\in\frac{1}{S_q}\mathbb{N}$ obeying
\begin{equation}\label{SqS1}
 S_q = \Big\lfloor \frac{t+1}{t^{\frac{q}{2-q}}+1} S_1 \Big\rfloor
\end{equation}
such that $(t+1)S_1 = (l^q+1)S_q$ and
\[
 \delta_{(l^q+1)S_q} < \frac{\rho(1-\kappa^4) + \kappa^2 \sqrt{4\rho+1}} { \rho (\kappa^2+1)^2 + \kappa^2 }
\]
with $\rho= \frac{1}{4} l^{q-2}$.

\end{theorem}
It can be verified that Theorem~\ref{thm:5} also holds for the condition (\ref{noisy_cond}). The equation (\ref{SqS1}) states that the $\ell_q$-analysis minimization with $q<1$ can do recovery in a wider range of cosparsity than the $\ell_1$-analysis minimization. For example, if $\delta_{5S_1}<\frac{2\sqrt{5}}{5}$, then the $\ell_{\frac{2}{3}}$-analysis minimization can recover a vector $\betaB$ with $||\DB\betaB||_0=S_{\frac{2}{3}}=\lfloor \frac{5}{3} S_1 \rfloor $.

\section{Iteratively Reweighted Method for $\ell_q$-analysis Minimization}
To solve the $\ell_q$-analysis minimization problem, we resort to the iteratively reweighted method. It has been demonstrated to be an effective approach for the $\ell_q$ norm related optimization problem; see \cite{GorodnitskyRao1997,ChartrandYin2008,Daubechiesabc2010}.

Note that $||\DB\betaB||_q^q$ has the following variational formulation:
\[
 ||\DB{\betaB}||_q^q = \sum_{i=1}^n (|\boldsymbol\DB_{i.} \betaB|^{\alpha})^{\frac{q}{\alpha}} = \min_{\boldsymbol{\eta} > \boldsymbol{0}} \Big\{ J_{\alpha} \triangleq \frac{q}{\alpha}  \sum_{i=1}^n \Big( \eta_i|\boldsymbol\DB_{i.} \betaB|^{\alpha} + \frac{\alpha-q}{q} \frac{1}{{\eta_i}^{\frac{q}{\alpha-q}}} \Big) \Big\}
\]
for $0<q\leq 1$ and  $\alpha\geq 1$. Its minimizer is attained at $\eta_i=1/|\DB_{i.} \betaB|^{\alpha-q}$, $i=1,\ldots,n$. However, when ${\betaB}$ is orthogonal to some $\DB_{i.}$, the weight vector $\boldsymbol{\eta}$ may include infinite components. To avoid an infinite weight, we add a smoothing term $q/\alpha \sum_{i=1}^n \eta_i\varepsilon^{\alpha}$ $(\varepsilon\geq 0$) to $J_{\alpha}$. Using the above variational formulation, we obtain an approximation of the problem (\ref{analysis_lq}) as
\begin{equation} \label{approx_lq}
 \min_{{\betaB},\boldsymbol{\eta}>\boldsymbol{0}} \frac{q}{\alpha} \sum_{i=1}^n  \Big[ \eta_i(|\boldsymbol\DB_{i.}{\betaB}|^{\alpha}+\varepsilon^{\alpha}) {+} \frac{\alpha{-}q}{q} \frac{1}{{\eta_i}^{\frac{q}{\alpha{-}q}}} \Big]   ~~ \textrm{ s.t. }~~ ||\yB-{\XB}{\betaB}||_2 \leq \epsilon  .
\end{equation}
By setting $\alpha=2$, we can apply the IRLS method to solve (\ref{approx_lq}). At the $k$-th iteration, we firstly compute the weight vector $\boldsymbol{\eta}$ via $\eta_i^{(k)} = ( |\boldsymbol{\DB}_{i.} {\betaB}^{(k-1)}|^{2} + {\varepsilon^{(k-1)}}^{2}  )^{q/2-1}$, with $\betaB$ fixed. Then update $\betaB^{(k)}$ with $\boldsymbol{\eta}$ fixed by solving 
\begin{equation}\label{cons_ls}
 {{\betaB}}^{(k)} = \argmin_{{\betaB}\in \mathbb{R}^d}  \frac{q}{2} \sum_{i=1}^n  \eta_i |\boldsymbol\DB_{i.}{\betaB}|^{2}   ~~ \textrm{ s.t. }~~ ||\yB-{\XB}{\betaB}||_2 \leq \epsilon  .
\end{equation}
Instead of (\ref{cons_ls}), we solve an equivalent alternative problem (with conjugate gradient) 
\begin{equation*}\label{rwl2}
 {{\betaB}}^{(k)} = \argmin_{{\betaB}\in \mathbb{R}^d} \Big\{\frac{1}{2}||\yB-{\XB}{\betaB}||_2^2 + \frac{\lambda q}{2} \sum_{i=1}^n \eta_i^{(k)} |\boldsymbol\DB_{i.}{\betaB}|^{2} \Big\}
\end{equation*}
for appropriate value $\lambda>0$ corresponding to $\epsilon$. In the implementation, the value of $\lambda$ is varied until the constraint in (\ref{cons_ls}) is satisfied. Lastly, we update $\varepsilon$ via $ \varepsilon^{(k)} = \min\{\varepsilon^{(k-1)}, \rho \cdot r(\DB{\betaB}^{(k)})_{l}\}~\textrm{ with } \rho\in(0,1]$, where $r(\DB{\betaB})_l$ denotes the $l$-th smallest element of the set $\{|\DB_{j.}{\betaB}| : j=1,\ldots,n\}$. In this paper, we call the above procedure as CoIRLq since it solves the $\ell_q$-analysis minimization problem. The pseudo-code of CoIRLq is given in the supplemental material.

\begin{remark}
 \rm Very recently, \citeauthor{Ochsabc2015} (\citeyear{Ochsabc2015}) propose a general iteratively reweighted algorithm for nonconvex optimization problem and show that it converges to a critical point. The CoIRLq algorithm is actually a special case of that conceptual iteratively reweighted method, and satisfies the assumed conditions.
 Hence the CoIRLq algorithm converges to a critical point of the problem (\ref{approx_lq}). Especially, when $\varepsilon\rightarrow 0$, the sequence $\{\betaB^{(k)}\}_{k\in\mathbb{N}}$ generated by the CoIRLq algorithm converges to a critical point of the problem (\ref{analysis_lq}).
\end{remark}

\section{Numerical Analysis}

In this section we compare the performance of the $\ell_q$-analysis minimization in the case $q<1$ and $q=1$ on cosparse vector recovery.
We generate the simulated datasets according to 
\[ \yB={\XB}{\betaB}+\wB, \]
where $\wB \sim N(\boldsymbol{0},\sigma\IB)$. The sampling matrix ${\XB}$ is drawn independently from the normal distribution with normalized columns. The analysis operator $\DB$ is constructed such that $\DB^T$is a random tight frame. To generate a $l$-cosparse vector ${\betaB}$, we first choose $l$ rows randomly from $\DB$ and form $\DB_{\Lambda}$.Then we generate a vector which lies in the null space of $\DB_{\Lambda}$. The recovery is deemed to be successful if the recovery relative error $||\hat{{\betaB}}-{\betaB}^{*}||_2/||{\betaB}^{*}||_2 \leq 1e-4$.

In the first experiment, we test the vector recovery capability of the CoIRLq method with $q=0.7$. We set $m=80,n=144,d=120,l=99,$ and $\sigma=0$. Figure 1 illustrates that the CoIRLq method recovers the original vector perfectly.
\begin{figure}[htb]
\centering
{
\includegraphics[width=0.4\linewidth]{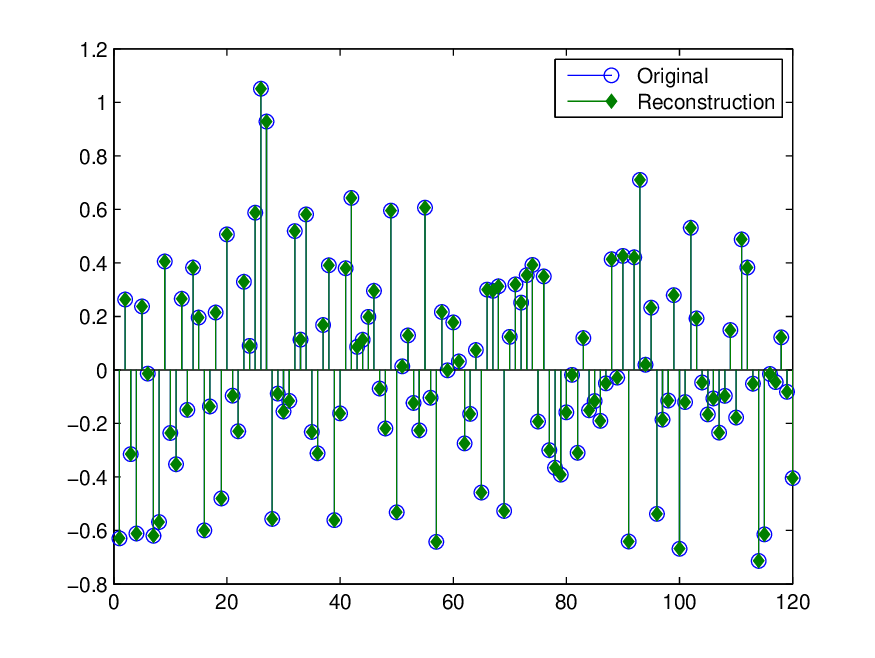}  \\
}
\caption{Cosparse vector recovery. }
\label{fig:1}
\end{figure}

In the second experiment, we test the CoIRLq method on a range of sample size and cosparsity with different $q$ in the noiseless case. Figure 2 reports the result with 100 repetitions on every dataset. We can see that the CoIRLq method with $q<1$ can achieve exact recovery with fewer samples and in a wider range of cosparsity than with $q=1$. 
Note that there is a drop of recovery probability where the cosparsity $l=118$ \footnote{When $l=120$, a zero vector is generated by our codes. So the recovery probability in cosparsity $l=120$ is zero.}. This is because  it is hard to algorithmically recover a vector residing in a subspace with a small dimension; please also refer to \cite{Namabc2011}. 
\begin{figure}[htb]
\center
\scriptsize
\center
    \begin{tabular}{c@{}c}
        \includegraphics[width=0.38\linewidth]{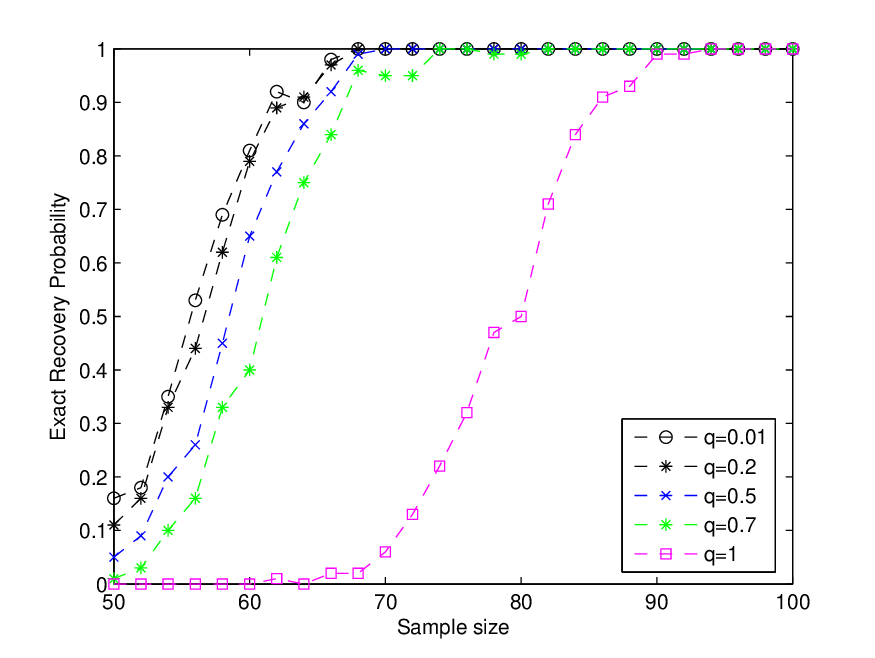}
        ~~~&~~~\includegraphics[width=0.38\linewidth]{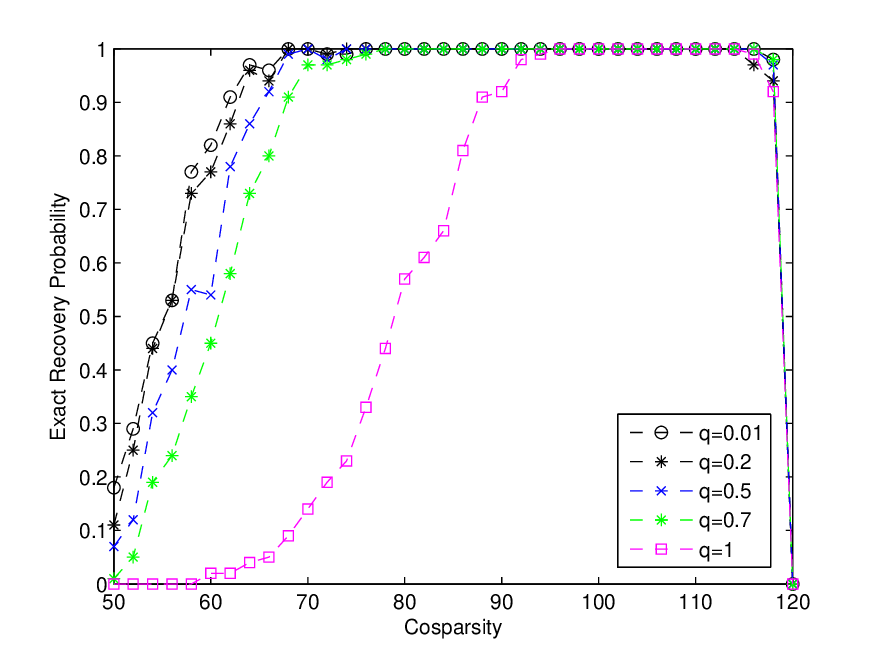}\\
        $n=144$, $d=120$, $l=99$ ~~~&~~~ $m=90$, $n=144$, $d=120$ 
    \end{tabular}
\caption{Exact recovery probability of the CoIRLq method.}
\label{fig:2}

\end{figure}


\section{Conclusion}

In this paper we have conducted the  theoretical analysis and developed the computational method, for the $\ell_q$-analysis minimization problem. Theoretically, we have established weaker conditions for exact recovery in the noiseless case and a tighter non-asymptotic upper bound of estimate error in the noisy case. In particular, we have presented a necessary and sufficient condition guaranteeing exact recovery. 
Additionally, we have shown that the nonconvex $\ell_q$-analysis optimization can do recovery with a lower sample complexity and in a wider range of cosparsity. Computationally, we have devised an iteratively reweighted method to solve the $\ell_q$-analysis optimization problem. 
Empirical results have illustrated that the nonconvex method outperforms its convex counterpart.

\section*{References}

\bibliography{mybibfile}

\end{document}